    \documentstyle[aps,pre,amstex,twocolumn,psfig]{revtex}

\begin{document}

\draft
\title{Amplitude measurements of Faraday waves\\}
\author{A.~Wernet$^1$, C.~Wagner$^1$, D.~Papathanassiou$^{2}$,  H.~W.~M\"uller$^{3}$ and K.~Knorr$^1$ }
\address{$^1$ Institut  f\"ur Technische Physik, Universit\"at
des Saarlandes, Postfach 151150, D-66041 Saarbr\"ucken, Germany\\
$^2$ Institut  f\"ur Theoretische Physik, Universit\"at des
Saarlandes, Postfach 151150, D-66041 Saarbr\"ucken, Germany\\
$^3$Max Planck Institut f\"ur Polymerforschung, Ackermannweg 10,
D-55128 Mainz, Germany } \maketitle

\begin{abstract}
A light reflection technique is used to measure quantitatively
the surface elevation of Faraday waves. The performed
measurements cover a wide parameter range of driving frequencies
and sample viscosities. In the capillary wave regime the
bifurcation diagrams exhibit a frequency independent scaling
proportional to the wavelength. We also provide numerical
simulations of the full Navier-Stokes equations, which are  in
quantitative agreement up to supercritical drive amplitudes of
$\varepsilon \simeq 20\%$. The validity of an existing
perturbation analysis is found to be limited to $\varepsilon <
2.5\%$.
\end{abstract}
%
\narrowtext
\section{Introduction}
A sound understanding of  hydrodynamic pattern forming systems is
based on a balanced interplay between experiment and theory,
 both analytical and
numerical. During the past this concept has led to considerable
progress, especially in case of  Rayleigh-B\'{e}nard convection (RBC)
or Taylor-Couette flow (TCF), where an amazing level of
quantitative agreement has been achieved. Another famous example
of
 pattern forming systems is the Faraday experiment: Surface waves on
a free liquid-air or liquid-liquid interface are excited by a
sinusoidal vibration of the fluid layer in vertical direction.
This system has the advantage of fairly short time scales in
combination with a very rich bifurcation behavior.  During the
last two decades the focus in this field  was on nonlinear pattern
selection
\cite{miles90,milner90,Christiansen92,binks97,kudrolli97},
secondary instabilities \cite{douady89}, the transition towards
chaos \cite{bosch93}, droplet ejection \cite{goodridge96} and
stokes drift \cite{ramshankar95}. However, due to the parametric
drive mechanism the mathematical description of the Faraday
experiment is more complicated as compared to RBC or TCF and the
quantitative understanding is less advanced yet. For instance, it
was not until recently that a rigorous linear stability theory had
been developed, valid for viscous fluids and realistic boundary
conditions\cite{kumar94}. Nowadays the predictions of the linear
theory and the experimental results agree within a few percent
\cite{bechhoefer95}. On the non-linear level, however, the
agreement between theory and experiment can at best be considered
qualitative: In the framework of a weakly non-linear perturbation
analysis different primary surface wave patterns with quadratic,
hexagonal etc. symmetry have been predicted \cite{zhang96}. Even
though most of them have been found in experiments
\cite{binks97,kudrolli97}, empiric and predicted phase diagram
reveal only a qualitatitve coincidence. Moreover, other experiments
operate with a two-frequency drive signal
 \cite{edwards94,kudrolli98}, or at very shallow filling heights
\cite{wagner00} or with complex fluids
\cite{raynal99,wagner99,lioubashevski99}. The resulting pattern
dynamics becomes more complex and exotic structures like
superlattices \cite{kudrolli98,wagner00,wagner99} or oscillons
\cite{lioubashevski99} appear. In some cases a qualitative
understanding based on symmetry arguments could be obtained
\cite{wagner00,silber98}.

With regard to this situation it comes as a surprise that a
systematic quantitative investigation of the system's major order
parameter, the surface elevation, has not been undertaken yet.
Such data is indispensable for a verification of any nonlinear
theory. It is the aim of the present work to provide extensive
experimental material in order to fill this gap. A measurement
technique is presented appropriate to quantify the surface
elevation of Faraday ripples. Our measurements cover a broad part
of the parameter space explored in recent experiments on pattern
selection in the Faraday experiment. In case of low viscosity
fluids our findings are expected to compare with the perturbation
analysis of Zhang and Vin\~{a}ls \cite{zhang96} -- at least at a
weak supercritical drive. Furthermore, a closer comparison yields
a reliable estimate of the validity range of this approximation.
Farther away from threshold of the instability, quantitative
theoretical predictions are not available yet. Here we provide a
numerical simulation by means of a finite difference scheme. This
computation allows for 2-dimensional solutions of the full
Navier-Stokes equations. They are used for a comparison with our
measurements on line patterns.

\section{The system}
We consider a fluid layer of thickness $h$ with a free surface
vibrated in vertical direction at a drive frequency $\Omega$. In
the frame of reference co-moving with the container the liquid is
subject to a modulated gravity acceleration ${\rm g}+a \,
\sin{(\Omega t)}$, where ${\rm g}$ is the gravitational
acceleration and $a$ the amplitude of the drive. The fluid,
considered as being incompressible, is characterized by  the
kinematic viscosity $\nu$, the density $\rho$, and the surface
tension $\sigma$. The hydrodynamic problem is governed by the
Navier-Stokes equation, in which the modulated gravity enters as
a parametric drive. The boundary conditions are free slip at the
fluid-air interface and no slip along the walls and the bottom of
the container. If $a$ exceeds a critical threshold
$a_c(\Omega,h,\nu,\rho,\sigma)$ the surface, being plane at the
beginning, undergoes the Faraday instability and  standing waves
with a wavenumber $k$ appear. Usually these waves organize
themselves in form of regular patterns of different possible
symmetries (lines, squares, hexagons ...). In containers with a
large lateral aspect ratio (the ratio between container dimension
to wavelength of the pattern) the pattern selection is geometry
independent and solely governed by the nonlinearities in the
equations of motion. In contrast, by changing the lateral
container extension one can manipulate the selection process. For
instance, by reducing one sidelength of a beforehand quadratic
container, a line pattern (oriented parallel to the shorter side)
can eventually be enforced in a situation where squares would
prevail otherwise. In the present paper extensive use will be
made of this geometrical selection feature.

Generically, the time dependence of the Faraday mode bifurcating
out of the undisturbed plane surface is subharmonic, i.e. the
surface oscillates at half the drive frequency $\omega=\Omega/2$.
The standing wave surface profile $\eta({\bf r})$ at onset of the
instability can be written in the form
\begin{equation}
\label{surf1}
 \eta_N({\bf r},t) = \frac{1}{2} \, \sum_{i=1}^{N} {(
    A_i e^{{\rm i} {\bf k}_i \cdot {\bf r}} } + c.c.)
\,  \sum_{n=-\infty}^{+\infty}  \zeta_n
    e^{{\rm i} n  \frac{\Omega}{2} t}.
\end{equation}
Here ${\bf r}=(x,y)$ abbreviates the horizontal coordinates. The
set of Fourier coefficients $\{\zeta_n\}$ determines the
subharmonic time dependence. Exactly at the onset of the
instability, $a=a_c$, one has $\zeta_n=0$ for even $n$, while the
odd coefficients are the components of the eigenvector related to
the linear stability problem.  The spatial modes are characterized
by the wave vectors ${\bf k}_i$, each carrying an individual
amplitude $A_i$. These quantities are  determined by the
nonlinearities of the problem and -- if appropriate -- also by the
container geometry. In principle the ${\bf k}_i$ can have any
length and orientation but usually they are supposed to be equally
spaced on the circle $|{\bf k}_i|=k$. Then the number $N$ of
participating modes determines the degree of rotational symmetry
of the pattern:
 $N=1$ corresponds to lines, $N=2$ to squares,
$N=3$ to hexagons or triangles, etc. Patterns with $N>3$  are no
longer translational symmetric, we refer to them as
quasi-periodic.

\section{Prework by other authors}
\subsection{Experimental}

There exist a considerable number of measuring techniques
appropriate to investigate the surface wave dynamics. They range
from contacting permittivity measurements \cite{atatuerk87} over
optical systems \cite{battezz70} including interferometry, radar
back scattering \cite{nicolas97}  up to x-ray absorption
techniques \cite{richter00}. Most of these procedures apply to
gravity waves, i.e. long wavelength surface waves. Each
 of them has its own limitations. Either the amplitude
 is recorded locally in space, or the detector is too slow to resolve
 the temporal spectrum.  One of the strongest restrictions for
 optical reflection techniques comes from the poor reflectivity of
 most fluid-air interfaces. Only
a few percent of the incident light is reflected and an even
smaller contribution is due to diffusive back-scattering. Since
the  latter is crucial for standard visualization techniques such
as holographic photography or triangulation\cite{perlin93}, these
methods often do not work properly. One might think of spraying
the fluid surface with tracer particles like club moss seeds but
this in turn affects the surface tension or the surface viscosity
in an uncontrollable manner. In a pair of very recent publications
\cite{Dabiri96,Roesgen98} the authors propose sophisticated
methods based on a colored illumination or on an array of
microlenses. A standard visualization technique used in several
Faraday experiments  is the light shadowgraphy
\cite{Christiansen92,binks97,ciliberto84}. A beam of parallel
light with a diameter comparable to the container size passes
through the fluid layer. The deformed standing wave surface
pattern with its heaps and hollows acting like an array of
collecting and diffusing lenses  is mapped onto a screen. However,
this technique is restricted to very small surface deformations
for which the profile can be recovered by ray tracing. Otherwise
caustics occur, which make this simple and effective method break
down.

An alternative technique introduced by Wright, Budakin and
Putterman \cite{Wright96} is based on intensity losses due to
diffusive scattering. These authors use Polystyrene colloids to
provide light scatterers within the fluid. The transversing light
intensity, weakly modulated by the local layer thickness, is
recorded by a high-sensitivity  CCD-camera to reconstruct the
surface profile. As an example, a snapshot is presented taken at a
rather strong drive amplitude within the turbulent regime.

To our knowledge, previous quantitative investigations on
near-onset Faraday waves are restricted to the low-frequency
regime, i.e. below $\Omega \simeq 30 {\rm Hz}$. Douady
\cite{Douady90} presents an investigation, where the deflection
of a thin laser beam directed onto a wave node is recorded. To
prevent the nodes from a gradual drift, Douady uses the ''rimfull
technique'' to fix the position of the pattern relative to the
container boundaries. That way the structure experiences a mode
discretization dictated by the container geometry. Data for the
elevation maxima and the linear relaxation time $\tau$ are
provided for a silicone oil at a viscosity of $10{\rm cS}$ and
drive frequencies between $20$ and $30 {\rm Hz}$.

Several other authors report amplitude measurements in small
aspect ratio experiments at driving frequencies between 5 and 10Hz
\cite{Virnig88,Henderson89,Craig95}. In this situation a
considerable mismatch between frequency and wavelength may occur.
The resulting detuning renders  the primary bifurcation
hysteretic. An additional problem in these experiments is the
moving contact line between the fluid and the container. It may
also generate a hysteresis or even induce an irregular time
dependent wave dynamics.

\subsection{Theoretical}
\subsubsection{Numerical}
\label{numerical} A couple of numerical simulations on Faraday
waves is worth mentioning in the present context. Zhang and
Vin\~{a}ls \cite{zhang98} reduce the original full 3-dimensional
hydrodynamic problem to a set of 2-dimensional non-local
equations formulated in terms of the lateral coordinates only.
This is the outcome of the so called ''quasi-potential
approximation'', which considers the bulk flow as being potential
(inviscid). Vortical flow contributions in the viscous boundary
layer beneath the surface are accounted for by an  effective
boundary condition. Clearly this analysis is only valid for deep
layers ($h \to \infty$) and restricted to the low dissipation
limit $\nu k^2 /\Omega \ll1$. The derived equations for the
surface elevation $\eta({\boldsymbol r},t)$ are numerically
integrated by a pseudo-spectral algorithm. The principal concern
of these simulations is to obtain the most preferred pattern and
to investigate its resistance against secondary instabilities.
Stationary patterns with different rotational symmetries have
been observed. In particular, drive frequencies at the transition
between gravity and capillary waves, $\Omega/2 \pi \simeq 30 {\rm
Hz}$, give rise to the most interesting structures:
Quasi-periodic pattern with a 10-fold rotational symmetry
\cite{binks97}. In the capillary regime squares or lines are
observed depending on the viscosity of the sample fluid.

A somewhat different numerical procedure
 has been used  by Schultz et al.\
\cite{schultz98} and Wright et al.\ \cite{wright00}. Both
investigations are based on the Euler equations for inviscid
fluids. Dissipation is accounted for by a phenomenological linear
damping term introduced afterwards.  The numerical procedures
used  are respectively a boundary-integral method and a
vortex-sheet technique. That way  the profile of very steep
2-dimensional Faraday waves is investigated. Their findings
comprise dimpled crests, the formation of a plumelike shape, or
the beginning of droplet ejection.

Obviously, all of the aforementioned numerical work  is
restricted to the low dissipation limit and to large layer
thicknesses. Indeed, we are not aware of systematic numerical
simulations for viscous fluids on the basis of the full
hydrodynamic equations and boundary conditions.

\subsubsection{Analytical}
\label{analyt}
 The first step towards a theoretical understanding
of pattern forming systems is a linear  stability analysis. This
gives access to the threshold amplitude $a_c$, the critical wave
number $k_c$, and the most unstable mode.  For mathematical
convenience it is  popular to assume laterally infinite
geometries. Although it is known for a long time that the
stability problem for Faraday waves can be approximately mapped to
a parametrically driven pendulum \cite{benjamin54} a rigorous
stability analysis for viscous fluids dates back until recently
\cite{kumar94}. Quantities evaluated by this method will be
referred to in the following as ''results of the exact stability
analysis''.

A nonlinear analysis suitable to predict the selected surface
pattern just above the primary instability has been presented by
Zhang and Vin\~{a}ls \cite{zhang96}. They start form their reduced
2-dimensional set of quasi-potential equations (see
Sec.~\ref{numerical}) and perform  a perturbation analysis for
small supercritical drive $\varepsilon=(a-a_c)/a_c \ll1$. The
analysis can be understood as a double expansion in $\varepsilon$
and in the dimensionless damping parameter $\gamma=\nu k^2/\Omega
\ll 1$. Since $h \to \infty$ is assumed, vortical flow
contributions coming from the viscous boundary layer along the
bottom of the container are ignored. The pattern wavelength $2
\pi/ k$ is approximated by the inviscid dispersion relation
\begin{equation}
\label{dispersion}
 \left ( \frac{\Omega}{2} \right )^2 = {\rm g} k + \frac{\sigma}{\rho}k^3 .
\end{equation}
In our experiments the damping parameter is $0.1<\gamma<0.2$ for
the low viscosity sample S1 and lies between $1.3$ and $1.7$ for
probe S2. Accordingly, the  wave number $k$ evaluated by
Eq.~(\ref{dispersion}) for S1 agrees within $1\%$ with the result
of the exact stability analysis but it is off by $20\%$ for S2. In
the framework of the weakly non-linear analysis the surface
profile is represented by Eq.~(\ref{surf1}) with $|k_i|=k_c$ and
$\zeta_n=0$ for $n$ even. The mode amplitudes $A_i$ are governed
by the set of the amplitude equations
\begin{equation}
\label{ampleqn1} \tau \partial_t A_i= \epsilon  A_i -
\sum_{j=1}^{N} \Gamma(\theta_{ij})|{A_j}|^2 A_i,
\end{equation}
where $\tau$ is the time constant of linear damping, also an
outcome of the linear analysis. $\Gamma(\theta)$ is the non-linear
coupling function, whose dependence on the sample specifications
and the drive frequency is analytically known. Evaluating $\Gamma$
at the angle increments $\theta_{ij}$ between two interacting
modes ${\bf k}_i$ and ${\bf k}_j$ yields the set of cubic coupling
coefficients, which governs the pattern selection process.  By
computing the stationary solutions of Eq.\ref{ampleqn1} in the
form $|A_i|={\cal R}_N$ with $i=1...n$ and
\begin{equation}
\label{squareroot} |A_i|={\cal
R}_N=\sqrt{\frac{\varepsilon}{\sum_{i=0}^{N-1} \Gamma(i \pi/N)}}
\end{equation}
at different symmetry indices $N$ one obtains the saturation
amplitude ${\cal R}_1$ for line patterns, ${\cal R}_2$ for
squares, etc. As outlined in Refs.~\cite{zhang96,mueller94} that
pattern with the smallest free energy
\begin{equation}
\label{lyap}
 {\cal F}_N=-\frac{1}{2}\, \epsilon^2
 \, \left \{ \sum_{l,j=1}^N \Gamma(\theta_{ij})|A_l|^2 |A_j|^2 \right \}
\end{equation}
is to be selected. Broadly speaking, in the capillary regime
investigated here, it is either squares (at low sample
viscosities) or lines (at higher viscosities).  Wave patterns with
a higher order rotational symmetry are not found since they
routinely posses a larger free energy. Since our optical
reflection technique works best with a 1-dimensional surface
modulation, we make use of the above mentioned geometrical
selection feature and perform all experiments in containers of
rectangular cross section. That way we enforce line patterns at
any investigated sample viscosity and under all operating
conditions. Under the assumption that the saturation amplitude for
lines is not appreciably influenced by the container geometry we
are able to compare our data with the line solution ($N=1$) of
Eq.~\ref{lyap}.
\begin{equation}
\label{surf2}
 \eta(x,t) = A \cos( { k_c}x )\,   \sin ( \frac{\Omega}{2} t).
\end{equation}
Here the saturation amplitude is given by

\begin{equation}\label{}
 A=\sqrt{\varepsilon/\Gamma},
 \end{equation}
where we have abbreviated $A_1$ by $A$ and $\Gamma(0^\circ)$ by
$\Gamma$.

\section{Experimental setup}
\begin{figure}
\psfig{file=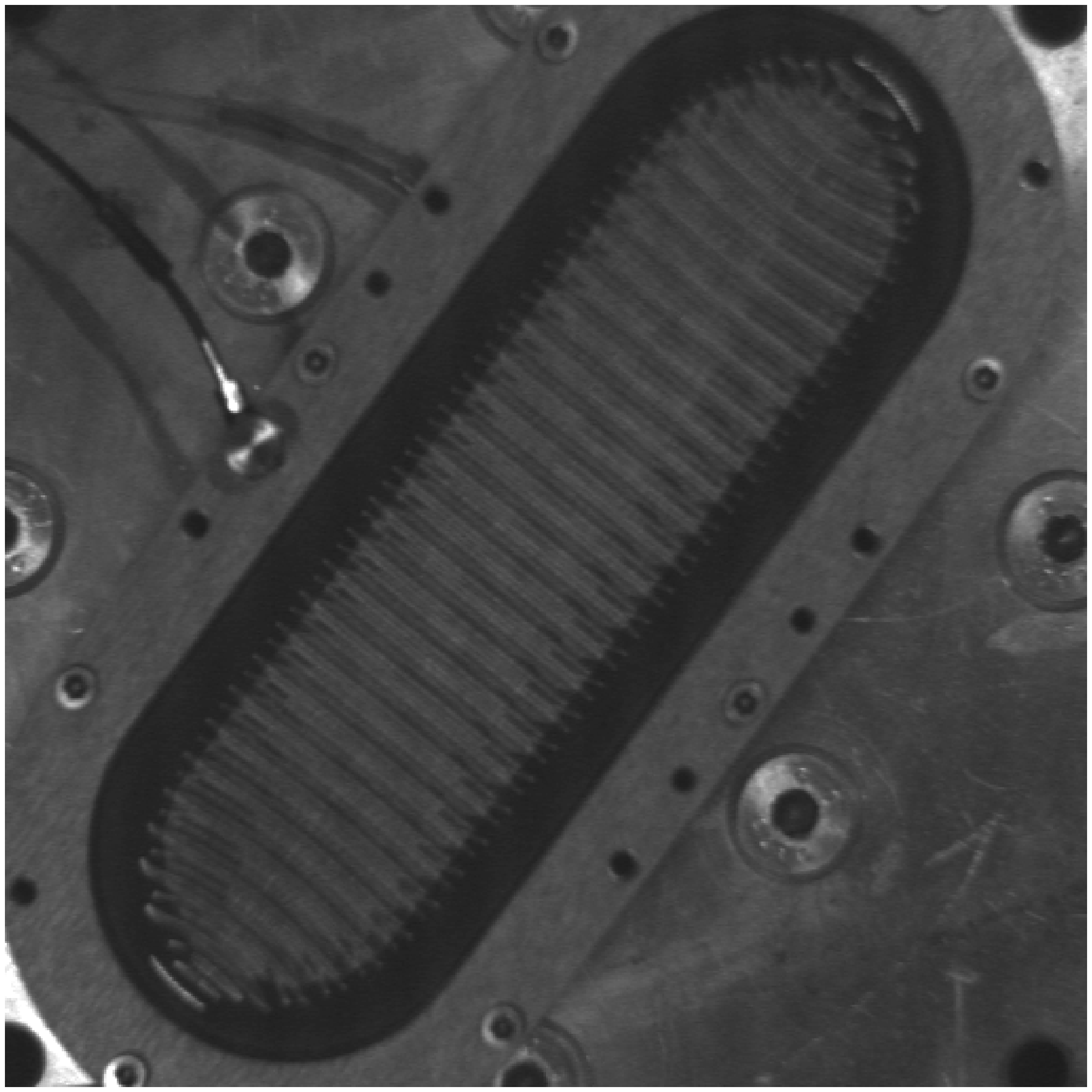,width=1.\columnwidth,angle=0} \caption[] {
Photograph of the container and a surface pattern of lines. The
screen for the optical deflection technique is dismantled and the
pattern is illuminated by a diffusive light source.  Due to the
rectangular geometry the lines are oriented perpendicular to the
longitudinal container axis (transverse mode). The deformations
appearing at the arched ends die out after a few pattern
wavelengths.  The accelerometer and some cabling are visible  on
the left  side of the container. The background of the picture is
the shaker armature. } \label{fig1}
\end{figure}

The surface wave excitation is accomplished by a vibration exciter
with a maximum force peak of $4670 {\rm N}$. Details are described
elsewhere \cite{wagner99}. The acceleration signal is computer
controlled, its amplitude and harmonicity are stabilized such that
fluctuations are smaller than $\pm0.2\%$.
 We use a container with a rectangular cross section (length $150 mm$, width $50 mm$)
in the form of a stadium (see Fig.~\ref{fig1}). This side length
ratio is sufficient to enforce Faraday patterns in form of lines
for all investigated sample fluids and under all  operating
conditions. To minimize disturbances originating from meniscus
waves, the form of the rim is designed as a ''soft beach'', where
the depth increases up to its maximum of $5{\rm mm}$ on a length
of $9{\rm mm}$ giving an inclination angle of $\simeq 34^\circ$.
 The curved sides of the
stadium also help to  suppress meniscus waves due to destructive
interference.   The vessel was covered by a glass plate to avoid
pollution, evaporation  and temperature fluctuations. The
temperature of the container (typically $T=30^\circ {\rm C}$) is
regulated with an accuracy of $\pm 0.1^\circ$. Two different {\em
Dow Corning 200} silicone oils are used as sample fluids. For a
complete specification see Tab.~I. The choice of the filling
height $h=3{\rm mm}$ is large enough to guarantee that the finite
depth correction to the dispersion relation (\ref{dispersion})
can be ignored, i.e. $\tanh(kh)>0.995$ for all measurements.

The knowledge of the wave number $k$ is   crucial for the
interpretation of the  elevation amplitude (see below). Therefore,
before each run of amplitude measurements we evaluate the wave
number $k$ of the pattern by photographing the free surface
illuminated by a diffuse light source (see Fig.~\ref{fig1}).

\begin{figure}
\psfig{file=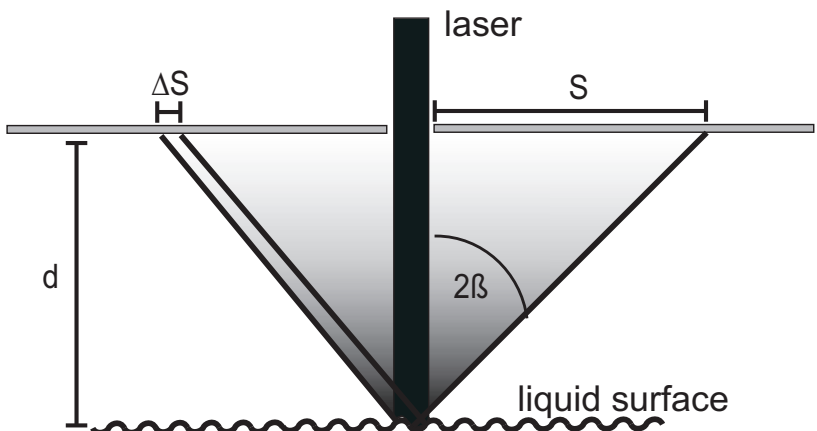,width=1.25\columnwidth,angle=0} \caption[]{
Sketch of the amplitude measurement technique. A laser beam with a
diameter comparable to the wavelength of the surface pattern
passes through a hole in the center of a diffusing screen and gets
reflected at the oscillating fluid surface. In case of a line
pattern the laser spreads out to a light streak of length $s$ on
the screen, which is recorded by a CCD camera.  Since the light
ray defining the tips of the streak may originate from two
neighboring wave nodes the length $s$ is flawed by a systematic
error $\Delta s$ of one wavelength $2 \pi/k$. }
 \label{fig2}
\end{figure}

To record surface wave amplitudes we use a laser beam directed
vertically onto the fluid surface (see Fig.~\ref{fig2}). The cross
section of the beam is widened to a diameter between $0.75$ and
$1.25$ of a pattern wavelength. The light beam reflected at the
standing wave surface pattern hits a diffusive screen mounted
above the liquid-air interface. The shape of the reflected light
pattern depends on the surface wave structure. In case of lines,
a bright light streak with sharp edges occurs on the screen. Its
length, oscillating with the frequency of the external drive, is
recorded by a CCD camera and digitized. The largest length during
an oscillation cycle yields the maximum surface slope $\partial
\eta(x)/\partial x |_{max}$. According to the geometry shown in
Fig.~\ref{fig2} one obtains
\begin{equation}
\label{methodeqn} \frac{\partial \eta}{\partial x}|_{{\rm
max}}=\tan(\beta) = \tan{(\frac{1}{2}\arctan{\frac{s}{d}})}.
\end{equation}
Deflection angles up to $\beta_{{\rm max}}=40^\circ$ have been
exploited. Minor effects due to the refraction of the light beam
by the glass cover are corrected for. As seen in Fig.~\ref{fig2}
the light rays marking the tips of the streak may originate from
two neighboring elevation nodes. This and other errors in $s$
together with the inaccuracies in $k$ and $d$ sum up to a
relative systematic error in $\partial \eta/\partial x |_{{\rm
max}}$ which does not exceed $10\%$ for dive amplitudes
$\varepsilon > 0.5\%$ (see Fig.~\ref{fig3}a). In case the surface
profile at the moment of maximum elevation is approximately
sinusoidal with wave number $k$ the amplitude $\eta|_{{\rm max}}$
can be deduced from (\ref{methodeqn}) via
\begin{equation}
\label{slopeeqn} \eta|_{{\rm max}}=\frac{1}{k} \, \frac{\partial
\eta}{\partial x}|_{{\rm max}}.
\end{equation}
The light pattern on the diffusive screen is recorded by a CCD
camera situated vertically above it. The pictures are evaluated by
a home made processing software. To that end each image is
binarized and the longest distance between any two points of the
light streak is extracted. An automatic adaption of the
binarization threshold is implemented to compensate for the
decreasing light intensity of the streak during an amplitude ramp.

The measurement of the bifurcation diagram $\partial \eta/\partial
x|_{{\rm max}}(\varepsilon)$ runs as follows: Starting at a drive
amplitude of $\varepsilon < 0.2\%$, the computer performs an
automatic ramp in steps of $\Delta \varepsilon = 0.065\%$ up to
the  maximum $\varepsilon_{{\rm max}}$. For sample S1 it is
$\varepsilon_{{\rm max}}=20\%$ but only $\varepsilon_{{\rm
max}}=2\%$ for the more viscous probe S2. The fluctuations of the
drive acceleration vary from $\pm0.05\%$ at low drive ($a \simeq
1 {\rm g}$) up to $\pm0.2\%$ at $a \simeq 20{\rm g}$. Between each
increment the ramp is suspended for a waiting period of $1$
minute to allow for the system to equilibrate to the new
situation. Then a series of $25$ snapshots of the light streak is
taken at regular intervals of $20 {\rm s}$. This yields the
average streak length and the statistical error as indicated by
the error bars in Fig.~\ref{fig4}b. After each upwards amplitude
ramp a second scan in opposite direction down to
$\varepsilon=-2\%$ is performed to check for a possible
hysteresis in the bifurcation diagram.

\section{Numerical simulations}
In this paper we present extracts from our numerical simulations
of the full nonlinear hydrodynamic problem adapted to  treat
2-dimensional Faraday patterns in form of lines. A sketch of the
implemented algorithm is given here, details will be presented
elsewhere \cite{dohwm}.

The line patterns are considered 2-dimensional in the
$x$-$z$-plane, the $y$-direction is ignored. The applied algorithm
is based on a popular marker-and-cell-method (MAC)
\cite{harlow,hirt} extensively used to simulate thermal convection
in fluids \cite{lueckeXX}. Therein the non-dimensionalized
incompressible evolution equations read as
\begin{eqnarray}
\partial_t u
 = -\partial_x u^2 - \partial_z (w u)
+ \frac{\Omega}{2\nu \, k^2} \, \bigl(\partial_{xx} u +
\partial_{zz} w \bigr)
- \partial_x p \enspace,  \\
\partial_t w =-\partial_x (u w) -\partial_z w^2
+ \frac{\Omega}{2\nu \, k^2} \, \bigl(\partial_{xx} u +
\partial_{zz} w \bigr)
- \partial_z p, \\
\label{incompress}
\partial_x u + \partial_z w = 0
\end{eqnarray}
with $u$ and $w$ being the $x$- and $z$-component of the velocity,
respectively and $p$ being the pressure. To non-dimensionalize we
have used the following unities: wave number $k$ for length,
$\Omega/2$ for time and $\rho \Omega^2/(4 k^2)$ for pressure. The
above set of equations is solved on a staggered grid for $u$,
$w$, $p$.  The time integration is carried out  by a forward time
step, while the diffusive space derivatives and the pressure
gradient is evaluated by central differences. The convective
terms are treated by a Donor-Cell-scheme.  With a damped
Jakobi-method the pressure is determined such that the
incompressibility condition (\ref{incompress}) is met. Since we
are not interested in surface profiles with breaking waves or
droplet ejection, $\eta(x,t)$ is a single valued function. Its
dynamics is determined by the kinematic boundary condition
\begin{equation}
\partial_t \eta = - u \left. \right|_{z=\eta} \, \partial_x \eta +
w \left. \right|_{z=\eta} \enspace.
\end{equation}
The dynamical boundary conditions
\begin{equation}
\label{tangential} {\boldsymbol t} \cdot \sigma^\prime \left.
\right|_{z=\eta} \cdot {\boldsymbol n}  = 0 \enspace,
\end{equation}
\begin{multline}
\label{normal} p \left. \right|_{z=\eta} - {\boldsymbol n} \cdot
\sigma^\prime \left. \right|_{z=\eta} \cdot {\boldsymbol n} =
\frac{4 \,\text{g} \, k}{\Omega^2} \, \bigl(1 + a \sin(2 t)
\bigr) \, \eta
\\
+ \frac{4 \, \sigma \, k^3}{\rho \, \Omega^2} \,
 ({\boldsymbol \nabla} \cdot {\boldsymbol n}) \enspace, \label{nornum}
\end{multline}
ensure the continuity of tangential stresses across the interface
and the discontinuity of normal stresses due to the finite
surface tension. Here $\sigma^\prime_{ij}=2 \frac{\nu
k^2}{\Omega}(\nabla_i u_j \nabla_j u_i)$ denotes the
dimensionless viscous stress tensor, $\boldsymbol n(x,t)$ is is
the surface normal vector in outward direction, and $\boldsymbol
t(x,t)$ the tangential vector perpendicular to it. Note that the
tangential condition is completely ignored in some previous free
surface algorithms \cite{ripple}. Here the implementation of
Eq.~(\ref{tangential}) is accomplished by approximating the
discretized interface line by either horizontal, vertical or
diagonal segments as suggested by Grieb \cite{grieb}. Besides the
surface boundary conditions we impose a no-slip condition at the
bottom. For the present simulations periodic boundary conditions
in lateral direction are used, even though our algorithm allows to
switch easily to a realistic no-slip situation. All simulations
are  performed by a mesh composed of $80$ cells per wavelength $2
\pi/k$ at a time step size of $0.001$.

We emphasize that the present integration method does not suffer
from the limitations of earlier algorithms (see
Sec.~\ref{numerical} above). In particular, since it is based on
the full Navier-Stokes equations, there is no restriction to the
weakly dissipative limit. Moreover, our algorithm allows to study
finite filling levels, even if the depth of the viscous boundary
layer compares to the layer thickness $h$.

\section{Results}

\subsection{Measurements at low viscosity }
\label{lowviscosity}
 By virtue of the rectangular container shape
the primary Faraday pattern  consists of lines oriented parallel
to the shorter sidewall. This is in contrast to control
experiments carried out in large aspect ratio vessels, where
squares are the selected planform under otherwise identical
conditions. At the onset of the instability defining
$\varepsilon=0$ the lines  occur first in local regions of the
surface. By increasing the drive up to $\varepsilon\approx 0.5\%$
the line pattern spreads out over the whole surface. This
non-ideal onset is due to the spatial inhomogeneity of the drive.
To rule out whether the finite longitudinal container dimension
gives rise to a mode discretization we scanned the drive frequency
between $\Omega/2 \pi=80 {\rm Hz}$ up to $100{\rm Hz}$ in steps of
$0.5 {\rm Hz}$. Thereby the number of waves fitting into the
container increased from $32$ to $37$ in a continuous manner. By
virtue of the ''beach like'' container rim no stepwise behavior
of the curves $k_c(\Omega)$ or $a_c(\Omega)$ could be detected.
Moreover, the measured onset amplitude and wave number always
agreed within $0.5\%$ with the theoretical prediction of the
exact linear analysis computed for a laterally infinite fluid
layer.

\begin{figure}
\psfig{file=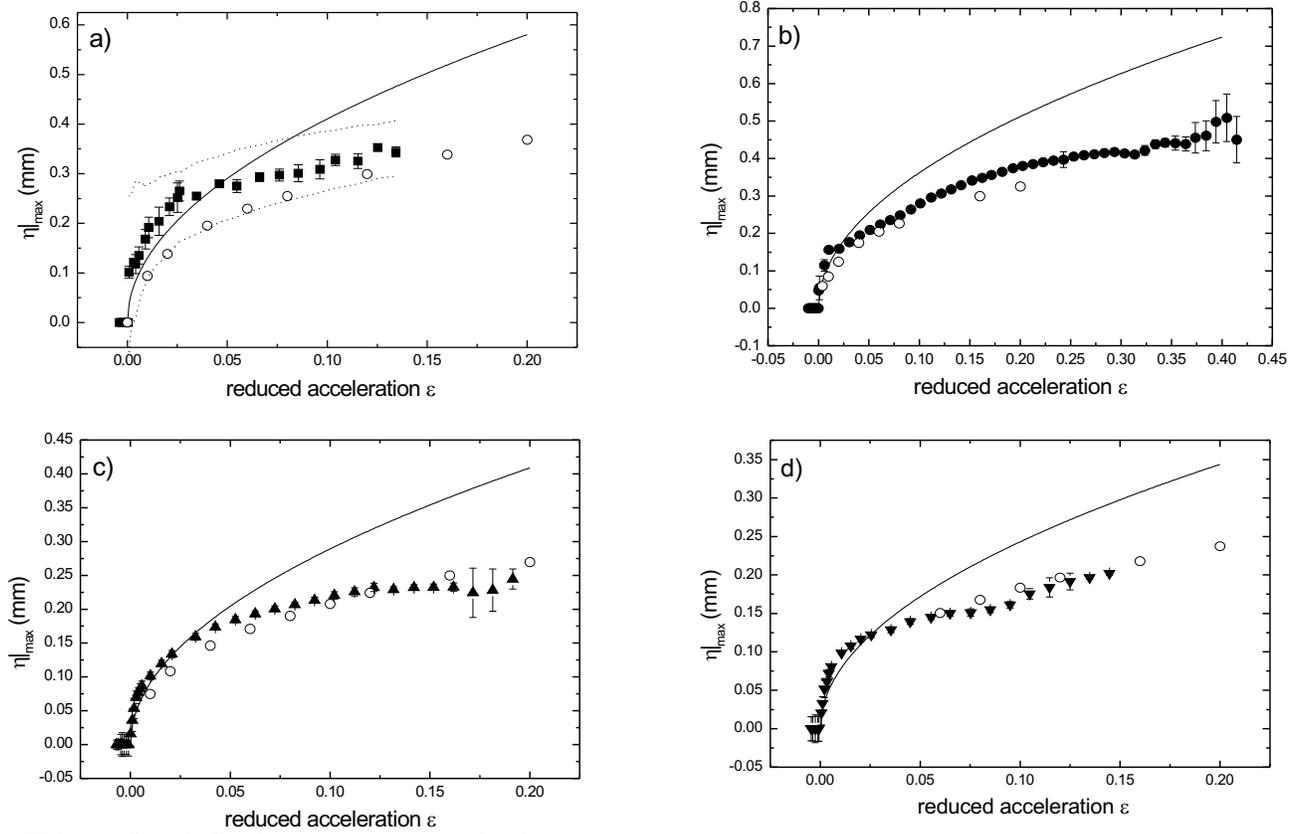,width=2.\columnwidth,angle=0} \caption[] {
Sample S1: Stationary wave amplitude  $\eta|_{{\rm max}}$ as a
function of the reduced acceleration $\varepsilon=a/a_c-1$ taken
at the drive frequencies (a): $\Omega/2\pi=60 {\rm Hz}$, (b): $80
{\rm Hz}$, (c): $120 {\rm Hz}$, (d) $160 {\rm Hz}$. The maximum
elevation $\eta|_{{\rm max}}$ is derived from the measured slope
via the relation $\eta|_{{\rm max}}=
\partial \eta/\partial x|_{{\rm max}}/k_c$. Solid symbols: experimental
data, solid line: result of the weakly nonlinear perturbation
analysis of Ref.~\cite{zhang96}, open symbols: results of the full
scale numerical simulation. The band between the dotted lines in
(a) indicates the systematic error of the measuring technique.
Error bars indicate statistical fluctuations as obtained by a
sequence of $25$ individual measurements. At small $\varepsilon$
the statistical fluctuation increase since the length of the light
streak is small. At large $\varepsilon$ the error increases
again  because of a beginning defect dynamics, which destroys the
stationarity and the coherence of the line pattern. } \label{fig3}
\end{figure}

A set of amplitude measurements performed at four different drive
frequencies is shown in Fig.~\ref{fig3}. Only the data obtained by
up-ramping the drive are plotted since the corresponding
down-ramps did not deviate significantly. There was no indication
for a hysteretic transition. The highest drive amplitude achieved
was $\varepsilon_{{\rm max}} \simeq 40\%$ (Fig.~\ref{fig3}b)
where the maximum surface elevation reaches about $0.5 {\rm mm}$,
i.e. already $16\%$ of the layer thickness. However in most runs
the amplitude scan had to be stopped at $\varepsilon_{{\rm max}}
\simeq 20\%$ due to an incipient defect dynamics. Over the whole
investigated drive amplitude range, $0 < \varepsilon <20\% \,
(40\%)$ no deviation of the non-linear wave number
$k(\varepsilon)$ from its onset value $k_c$ could be detected.

\begin{figure}
\psfig{file=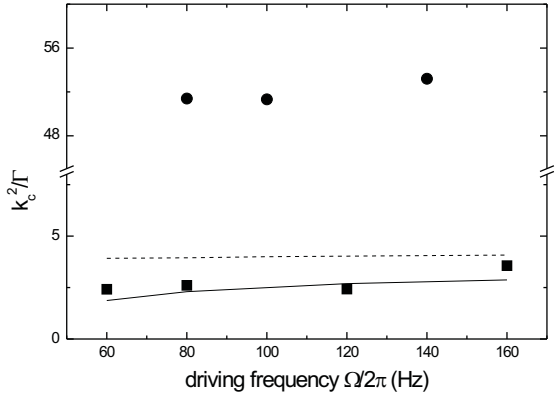,width=1.\columnwidth,angle=0} \caption[] {
 The coupling coefficient
$1/\Gamma$ (scaled by $k_c^2$) obtained from sample S1 (S2). The
squares (circles) denote fitted values for $\Gamma$ as compiled
from the data of Fig.~\ref{fig3} (Fig.\ref{fig6}). The solid
(dashed) line is the prediction of the perturbation theory
\cite{zhang96}.}
 \label{fig4}
\end{figure}

\begin{figure}
\psfig{file=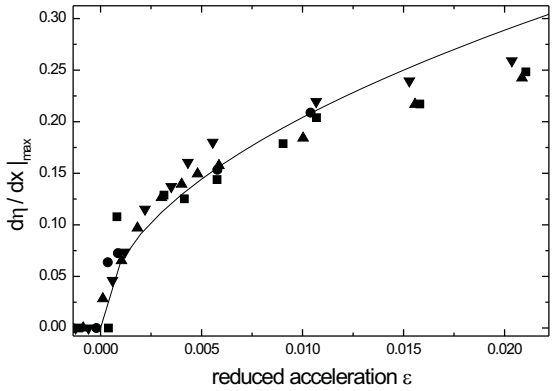,width=1.\columnwidth,angle=0} \caption[]
{Sample S1: Bifurcation diagram for the maximum surface slope
$\partial \eta/\partial x|_{{\rm max}}$. Symbols and parameters
as in Fig.~\ref{fig3}. The solid line is the asymptotics given by
Eq.~\ref{scaling} for $\Omega \to \infty$.  } \label{fig5}
\end{figure}

The bifurcation diagrams $\eta|_{{\rm max}}(\varepsilon)$ as
shown in Fig.~\ref{fig3} demonstrate that a square-root like
increase according to the theoretical prediction $\eta|_{{\rm
max}}=\sqrt{\varepsilon/\Gamma}$ (solid line in Fig.~\ref{fig3})
is limited  to rather small drive amplitudes below $\varepsilon
\simeq 2.5\%$.
 For a closer quantification we used the
experimental data at $0<\varepsilon < 2.5\%$ and fitted the
coefficient $\Gamma$. Good agreement with the prediction of Zhang
and Vin\~{a}ls  (see squares and solid line in Fig.\ref{fig4}) is
observed. It is interesting and to our knowledge unmentioned yet
that the analytical expression for $\Gamma/k_c^2$ \cite{zhang96}
becomes $\Omega$-independent at large drive frequencies according
to
\begin{equation}
\label{scaling} \Gamma/k^2 \rightarrow 3.84 + 2^{-\frac{2}{3}} \,
\frac{11}{240}\,
 \nu^{-2} \, (\frac{\sigma}{\rho})^{4/3} \,
\Omega^{-\frac{2}{3}} ,
\end{equation}
 It can be seen from
Fig.~\ref{fig4} that this asymptotics applies fairly well already
at drive frequencies beyond $\Omega/2\pi =60 {\rm Hz}$. As a
consequence of Eq.~(\ref{scaling}) the surface slope $\partial
\eta/\partial x|_{{\rm max}}$ should be asymptotically
independent of the drive frequency. This is tested by
Fig.~\ref{fig5}, where the surface slopes associated with
different drive frequencies approximately collapse on  a common
master curve. This scaling persists even for $\varepsilon >
2.5\%$, i.e. beyond the validity range of the perturbation
analysis.

Let us briefly discuss possible sources for discrepancies between
pertubation analysis and experiment. As mentioned above, the
appearance of line patterns aligned parallel to the shorter
container side (transverse mode) is enforced by the rectangular
container geometry. The result of this finite geometry effect may
be twofold, linear and non-linear. The linear one is negligible as
can be seen from the following argument: The container sidewalls
provide a damping offset to both, the longitudinal and the
transverse line mode. However, since the distance between the
longer sidewall pair is smaller the threshold shift for the
longitudinal mode is enhanced. Indeed, we even find the threshold
of the transverse mode almost unaffected: The empiric Faraday
onset $a_c$ agrees within $0.5\%$ with the prediction of the
exact stability analysis computed for a laterally infinite
container.
 Clearly, this linear reasoning does not imply that the saturated nonlinear
pattern amplitude remains unchanged too. However, estimating the
geometry effect upon the coefficient $\Gamma$ is a difficult task:
To that end one had to redo the perturbation analysis in terms of
the container eigenmodes instead of the simpler plane waves.

A second cause for deviations to the experimental results is the
restriction of the perturbation analysis to a pure sinusoidal time
dependence as given by Eq.~(\ref{surf2}). However, as outlined in
Sec.~\ref{analyt} the actual frequency spectrum of the Faraday
mode is not monochromatic. In the investigated region the error in
$a_c$ resulting from this approximation lies between $10\%$ and
$20\$$.

For a quantitative comparison with the experimental data at
elevated drive amplitudes  we refer to the numerical simulations
as described in Sec.~\ref{numerical} and indicated in
Fig.~\ref{fig3} by the open symbols. Up to elevated drive
amplitudes of $\varepsilon_{{\rm max}}\simeq 20\%$ we observe
good agreement between simulation  and experiment. The
deviation is nowhere worse than $15\%$ but considerably better at
the higher drive amplitudes, where the systematic error of the
detection method is smallest (see Figs.~\ref{fig3}c,d). Beyond
predicting the local elevation maximum the simulations also
provide access to the spatial anharmonicity of the surface
elevation. We find that the spatial harmonics $2k$, $3k$ ...
contribute to the Fourier spectrum by less than $4\%$. This
justifies a posteriori to equate the elevation amplitude
$\eta|_{{\rm max}}$ with the ratio $\partial \eta/\partial
x|_{{\rm max}} /k_c$ as used to produce Fig.~\ref{fig3}.

\subsection{Measurements at high viscosity}
\begin{figure}
\psfig{file=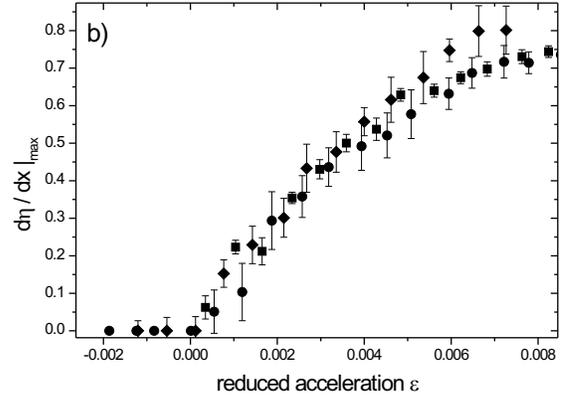,width=1.\columnwidth,angle=0} \caption[]
{Sample S2 at $T=30^\circ {\rm C}$: Stationary wave slope
$\partial \eta /
\partial x|_{{\rm max}}$ at the drive frequencies $\Omega/2\pi=80 {\rm Hz}$
(squares),  $100 {\rm Hz}$ (circles), $140 {\rm Hz}$ (diamonds).
} \label{fig6}
\end{figure}

Following the same procedure as in the previous section we
performed a set of amplitude measurements on the more viscous
fluid sample S2 (see Fig.~\ref{fig6}) at a temperature of
$30^\circ {\rm C}$. Again the selected pattern consists of
parallel transverse lines. However, unlike S1, this is the
preferred planform in large aspect ratio containers too, being  a
result of the elevated viscosity of probe S2. Consequently the
rectangular container geometry just determines the orientation of
the lines rather than altering the selected pattern.

Also in contrast to S1 the more viscous probe S2 exhibits a wave
amplitude, which grows rapidly with the driving acceleration
(Fig.~\ref{fig6}). The maximum deflection angle of $\beta_{{\rm
max}} \simeq 40^\circ$ is  reached at $\varepsilon < 1\%$.
Clearly, at those small drive  all bifurcation diagrams show
fairly well a square-root like increase according to
Eq.~(\ref{squareroot}). However, the coefficient $\Gamma$ as
compiled from the data at $0<\varepsilon < 0.8\%$  is an order of
magnitude larger than for S1 (circles in Fig.~\ref{fig4}).
Moreover, $\Gamma$ deviates substantially from the prediction of
the perturbation analysis (dashed line in Fig.~\ref{fig4}). But
the latter observation is not surprising since both the low
damping approximation as well as the infinite depth assumption
are violated for S2: Note that $1.3 < \gamma < 1.7 $ and the
depth of the viscous boundary layer is $0.5 {\rm mm}$, i.e. $20\%$
of the layer thickness. Although the perturbation analysis does
not quantitatively apply here, we recover the same
$\Omega$-independent scaling of  the bifurcation diagram. The
data for the slope $\partial \eta / \partial x|_{{\rm max}}
(\varepsilon)$ collapse again on a master curve (see
Fig.~\ref{fig6}) the same way as they did for S1 in
Fig.~\ref{fig5}.
\subsection{Viscosity dependent measurements}
\begin{figure}
\psfig{file=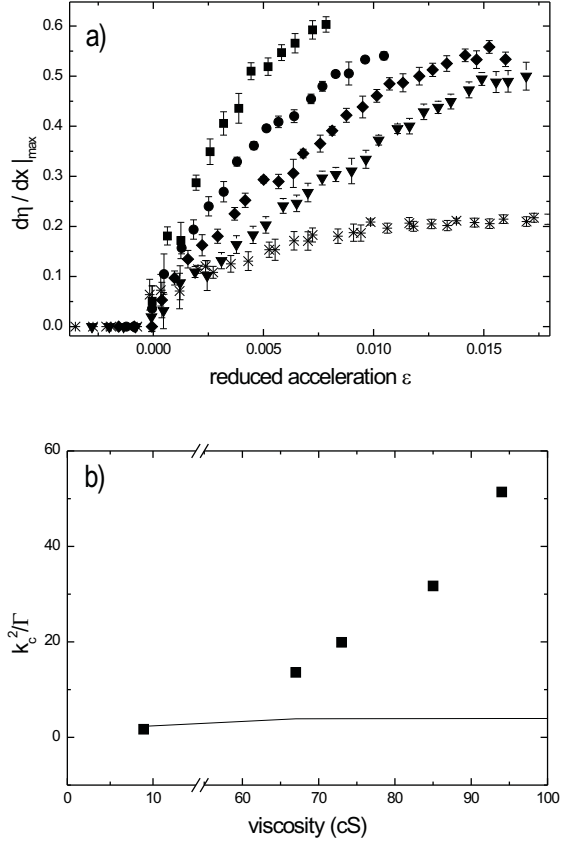,width=1.\columnwidth,angle=0} \caption[]
{Sample $S2$ at different temperatures: (a) Bifurcation diagram
 $\partial \eta
/
\partial x|_{{\rm max}}$  at the drive frequency $\Omega/2
\pi=80 {\rm Hz}$. The viscosities are
 $\nu=94 {\rm cS}$ (squares), $85 {\rm cS}$ (circles),
 $73 {\rm cS}$ (diamonds),  $63 {\rm cS}$ (down triangles), $8.35 {\rm cS}$ (stars).
 (b) The
coupling coefficient $1/\Gamma$ (scaled by  $k_c ^2$) as a
function of the viscosity $\nu$. The symbols show the coefficient
estimated by fitting the data of (a) for $\varepsilon < 1\%$. The
prediction of the perturbation analysis (solid line) becomes
unreliable for viscosities above  $\nu\simeq50 {\rm cS}$. }
 \label{fig7}
\end{figure}

A final set of measurements is devoted to the viscosity dependence
of the surface elevation.  Fig.~\ref{fig7}a shows a set of
bifurcation diagrams obtained with S2 at different viscosities
(see Tab.~\ref{tab1}). This is accomplished by varying the
temperature of the probe. These measurements are performed at the
drive frequency $\Omega/2 \pi=80 {\rm Hz}$. In agreement with our
previous observations the surface elevation steeply rises as the
viscosity increases. Fig.~\ref{fig7}b shows the dependence of the
coefficient $\Gamma$ on the viscosity as fitted from the
experimental data. By comparison with the result of the weakly
nonlinear analysis (solid line in Fig.~\ref{fig7}b) we conclude
that the small damping approximation $\gamma \ll 1$ holds at best
up to viscosities of $\nu\simeq 50 {\rm cS}$.

\section{Conclusions}
We have presented a series of systematic amplitude measurements
for stationary Faraday surface waves. The investigation is
accomplished by a laser beam reflected at the oscillating surface.
To facilitate the interpretation of the data the measurements are
performed on line patterns, which are enforced by the rectangular
container geometry. Due to the soft ''beach like'' boundary
conditions a mode discretization is avoided. Bifurcation diagrams
of the maximum surface deflection vs. the drive amplitude are
systematically recorded over a wide parameter range of drive
frequency and sample viscosity.

The experimental data reveal that the perturbation analysis of
Zhang and Vin\~{a}ls \cite{zhang96} applies quantitatively to
fluids with a viscosity of less than $\simeq 50 {\rm cS}$ and to
very small drive amplitudes of not more than $2.5\%$. Moreover, we
observe that the surface slope scales almost independently of the
drive frequency. This finding is also supported by the analytical
expression for the nonlinear coupling coefficient $\Gamma$ as
derived in the framework of the perturbation theory. Qualitatively
this scaling behavior persists even up to drive amplitude of
$\varepsilon_{{\rm max}} \simeq 20\%$, i.e. at operating
conditions, where the perturbation theory is no longer
applicable.

For a quantitative comparison of our data at elevated drive
amplitudes we provide a numerical simulation of Faraday waves on
the basis of the full Navier-Stokes equation. This new algorithm
does not suffer from the standard restrictions of the low
dissipation limit and large filling thicknesses as used  by other
previous simulations. Good quantitative agreement with the
empiric data is found up to the highest investigated drive
amplitudes of $\varepsilon_{{\rm max}} \simeq 20\%$.


{\it Acknowledgements\/} ---  We thank J.~Albers for his support.
This work is subsidized by Deutsche Forschungsgemeinschaft.

 

\clearpage


\clearpage

%
\clearpage


\clearpage
\begin{table}
\begin{tabular}{|c|c|c|c|c|c|c|c|}
  Sample & $T$ & $\Omega/2\pi$ & $\rho$ & $\nu$ & $\sigma$ & $a_c$ & $k_c$ \\
    & $[^\circ {\rm C}]$ & $[{\rm Hz}]$ & $[{\rm kg/m^3}]$ &
    $[{\rm cS}]$ & $[10^{-3}{\rm N/m}]$ & $[{\rm g}]$ & $[{\rm m^{-1}}]$
    \\ \hline
  $S1$ & 30 & 60 & 934 & 8.35 & 20.1  & 1.23 & 1060 \\
  $ $  &    & 80 &     &      &       & 1.91 & 1340 \\
  $ $  &    & 120&     &      &       & 3.85 & 1815 \\
  $ $  &    & 160&     &      &       & 6.2  & 2230 \\ \hline
  $S2$ & 30 & 80 &955.4& 94   & 20.55 & 15.1 & 1080 \\
  $ $  &    & 100&     &      &       & 21.7 & 1230 \\
  $ $  &    & 140&     &      &       & 37   & 1450 \\ \hline
  $S2$ & 35 & 80 &950.9&  85  & 20.2  & 14   & 1120 \\
  $ $  & 45 &    &941.9& 73   & 19.5  & 12.5 & 1180 \\
  $ $  & 50 &    &937.5& 67   & 19.15 & 11.7 & 1205 \\ \hline


\end{tabular}
 \label{tab1}
\caption{Fluid specifications at different temperatures for the
low (high) viscosity sample S1 (S2)}
\end{table}
\end{document}